\documentclass[mathleft]{an}
\usepackage{graphicx}
\usepackage{times}
\begin{document}
\sloppy

\newcommand{\pd}{\partial}
\def\onethird{{\textstyle{1\over3}}}
\def\onehalf{{\textstyle{1\over2}}}

\title{Turbulent viscosity and $\Lambda$-effect from numerical turbulence models}

\author{P. J. K\"apyl\"a\inst{1,2}\fnmsep\thanks{Corresponding author:
  \email{pkapyla@nordita.org}\newline} \and A. Brandenburg \inst{1}}

\titlerunning{Turbulent viscosity and $\Lambda$-effect from numerical
  turbulence models.}

\authorrunning{P.J. K\"apyl\"a \& A. Brandenburg}

\institute{ 
NORDITA, AlbaNova University Center, Roslagstullsbacken 23, SE-10691 
Stockholm, Sweden
\and
Observatory, University of Helsinki, PO BOX 14, FI-00014 University of 
Helsinki, Finland}

\date{Received $<$date$>$; 
accepted $<$date$>$;
published online $<$date$>$}

\keywords{Sun: rotation - hydrodynamics (HD)}

\abstract{Homogeneous anisotropic turbulence simulations are used
  to determine off-diagonal components of the Reynolds stress tensor
  and its parameterization in terms of
  turbulent viscosity and $\Lambda$-effect.
  The turbulence is forced in an anisotropic fashion by enhancing
  the strength of the forcing in the vertical direction.
  The Coriolis force is included with a rotation axis inclined
  relative to the vertical direction.
  The system studied here is
  significantly simpler than that of turbulent stratified convection
  which has often been used to study Reynolds stresses. Certain puzzling
  features of the results for convection, such as sign
  changes or highly concentrated latitude distributions, are not
  present in the simpler system considered here.}

\maketitle

\section{Introduction}
The Reynolds stress, described by the correlation of fluctuating
velocity components, $Q_{ij} = \overline{u_i u_j}$, is one of the most
important generators of differential rotation in stars (R\"udiger
\cite{Ruediger1989}). These stresses have been studied
with the help of 3D convection simulations (e.g.\ Pulkkinen et
al.\ \cite{Pulkkinenea1993}; Chan \cite{Chan2001}; K\"apyl\"a et
al.\ \cite{Kaepylaeea2004}; R\"udiger et
al.\ \cite{Ruedigerea2005}). These results have revealed some
surprising features such as the peaking of the horizontal stress $Q_{xy}$
very close to the equator, and a positive (outward) flux for rapid rotation.
Both of
these results are at odds with theoretical considerations (Kitchatinov
\& R\"udiger \cite{KitcRued1993}). Furthermore, disentangling of the
diffusive (turbulent viscosity) and non-diffusive ($\Lambda$-effect)
parts of the stress is difficult from convection simulations.

Here, we present preliminary results from anisotropic homogeneous, isothermal,
non-stratified turbulence simulations in which diffusive and
non-diffusive effects can be
studied separately. Imposing a linear shear flow on top of
isotropically driven turbulence allows the study of turbulent
viscosity without $\Lambda$-effect.
On the other hand, using a special form of forcing,
anisotropic homogeneous turbulence can be
generated. Rotation is added to study the
$\Lambda$-effect. A simple analytical closure model, based on the
minimal tau-approximation (hereafter MTA, see e.g. Blackman \& Field
\cite{BlackField2002}; Brandenburg et al.\ \cite{Brandea2004}), is used
to compare with simulations in the cases with
rotation.

\section{The models} 
\label{sec:model}
In the 3D simulations we solve the set of equations 
\begin{eqnarray}
\frac{\mathrm{D} \ln \rho}{\mathrm{D} t} &=& -\vec{\nabla} \cdot \vec{u}\;,\\
\frac{\mathrm{D} \vec{u}}{\mathrm{D} t} &=& -c_{\rm s}^2 \vec{\nabla} \ln \rho - 2\, \vec{\Omega} \times \vec{u} + \vec{f}_{\rm force} + \vec{f}_{\rm visc}\;,
\end{eqnarray}
using an isothermal equation of state characterized by sound speed
$c_{\rm s}$ in a fully periodic cube of volume $(2\pi)^3$. Here,
$\mathrm{D}/\mathrm{D} t = \partial/\partial t - \vec{u} \cdot
\vec{\nabla}$ is the advective derivative, $\rho$ is the density,
$\vec{u}$ is the velocity, and $\vec{\Omega} = \Omega_0 (-\sin \theta,
0, \cos \theta)^T$ is the rotation vector. By virtue of the periodic
boundaries mass is conserved and the volume averaged density has a
constant value of $\overline{\rho} = \rho_0$. In the present study we
use an anisotropic forcing function in Fourier space according to
\begin{eqnarray}
f_{i}^{\rm (force)} = (f_0\delta_{ij} + f_1 \cos^2\!\Theta_{\vec{k}}\,
\hat{z}_i \hat{z}_j) f_i^{\rm (iso)}\;,
\end{eqnarray}
where $f_0$ is the amplitude of the isotropic part and $f_1$ the
anisotropic one, $\Theta_{\vec{k}}$ is the angle between the vertical direction
and the wave vector $\vec{k}$, and $\hat{\vec{z}}$ is the unit vector
in the vertical direction. Details of the isotropic part of the
forcing are given, e.g., in Brandenburg et al.\ (\cite{Brandea2004}).

The viscous force is given by
\begin{eqnarray}
\vec{f}_{\rm visc} = \nu \Big( \nabla^2 \vec{u} + \onethird \vec{\nabla} \vec{\nabla} \cdot \vec{u} + 2\, \vec{\mathsf{S}} \cdot \vec{\nabla} \ln \rho \Big)\;,
\end{eqnarray}
where $\nu$ is the viscosity and
\begin{eqnarray}
\mathsf{S}_{ij} = \onehalf \bigg(\frac{\pd u_i}{\pd x_j} + \frac{\pd u_j}{\pd x_i} \bigg) - \onethird \delta_{ij} \frac{\pd u_k}{\pd x_k}\;,
\end{eqnarray}
is the rate of strain tensor. The simulations were made with the
{\sc Pencil-Code}\footnote{\texttt{http://www.nordita.org/software/pencil-code/}}.

In the minimal tau-approximation one solves for the time derivative of a
quantity instead of the quantity itself, i.e. in the present case
\begin{eqnarray}
  \dot{Q}_{ij} = \overline{\dot{u}_i u_j} + \overline{u_i \dot{u}_j}\;, \label{equ_dtQij}
\end{eqnarray}
where dots denote time derivatives and the overbars volume
averages. Inserting the Navier--Stokes equations into
Eq.~(\ref{equ_dtQij}) and assuming a high Reynolds number so that the
viscous terms can be neglected, one arrives at
\begin{eqnarray}
\dot{Q}_{ij} &=& -2\,\varepsilon_{jkl} \Omega_k Q_{il} -2\,\varepsilon_{ikl} \Omega_k Q_{jl} + \nonumber \\
&& \hspace{3cm} \overline{u_i f_j} + \overline{u_j f_i} + T_{ij}\;,
\end{eqnarray}
where $T_{ij}$ denotes the triple correlations. In the present case
we assume that the pressure terms are subsumed in the
triple correlations, which is reasonable for calculating the
$\Lambda$-effect, but not valid for calculating turbulent viscosity.
The basic assumption of MTA is that the
triple correlations can be presented in terms of the quadratic ones
\begin{eqnarray}
T_{ij} = - \tau^{-1} Q_{ij}\;,
\end{eqnarray}
where $\tau$ is a relaxation time.
In the statistically steady state without rotation we have
\begin{eqnarray}
Q_{ij}=Q_{ij}^{(0)}\equiv
\tau\left(\overline{u_i f_j} + \overline{u_j f_i}\right)\;,
\end{eqnarray}
which allows us to express the forcing in terms of the Reynolds tensor
for the non-rotating case (denoted by the superscript zero). We employ
the same amplitudes as those found in the 3D
simulations. The only free parameter in the MTA-model is then the
Strouhal number
\begin{eqnarray}
{\rm St} = \tau u_{\rm rms} k_{\rm f}\;, 
\end{eqnarray}
where $u_{\rm rms}$ is the rms-value of the fluctuating component of
the velocity and $k_{\rm f} \approx 5$ is the mean forcing wave number. In
the present study we use ${\rm St} = 2$ which reproduces the same
trend as a function of rotation as the 3D simulations (see
Fig.~\ref{fig:lambda}).

\begin{figure}
\resizebox{\hsize}{!}
{\includegraphics{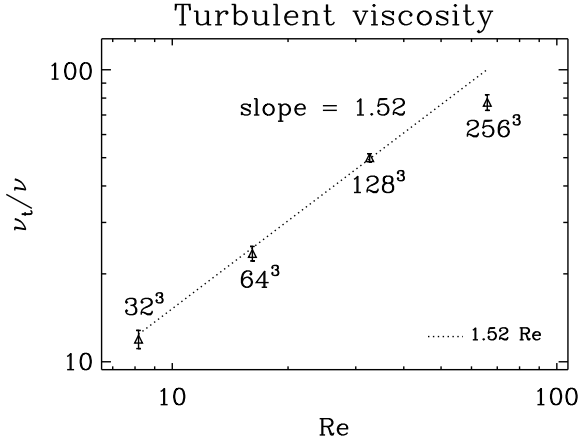}}
\caption{The ratio of turbulent to molecular viscosity, $\nu_{\rm
    t}/\nu$, as a function of the Reynolds number. The resolution used
  is denoted alongside each point. The error bars are estimated from a
  modified mean error of the mean; see, e.g. Eq.~(30) of K\"apyl\"a et
  al.\ (\cite{Kaepylaeea2004}).}
\label{fig:nut}
\end{figure}

\section{Results}
\label{sec:results}

\subsection{Turbulent viscosity}
In order to study the turbulent viscosity, a large scale linear shear
flow $\vec{U} = (0, Sx, 0)$ is imposed upon the system. In this system
the homogeneity of the turbulence is preserved by using the shearing
box approximation (e.g. Hawley et al.\ \cite{Hawleyea1995}). The
turbulent viscosity can now be computed from
\begin{eqnarray}
\nu_{\rm t} = - 2\,Q_{xy}/S\;.
\end{eqnarray}
where $S = -0.1$ is used in the present study.
Fig.~\ref{fig:nut} shows $\nu_{\rm t}/\nu$ as a function of
Reynolds number,
\begin{eqnarray}
{\rm Re} = \frac{u_{\rm rms}}{\nu k_{\rm f}}\;.
\end{eqnarray}
The absolute value of the turbulent viscosity stays almost constant as
a function of Reynolds number and thus the ratio $\nu_{\rm t}/\nu$
increases almost linearly as a function of ${\rm Re}$ except for the
largest Reynolds number run. Higher resolution simulations are needed in
order to clarify the behaviour in the high Reynolds number regime.

\subsection{$\Lambda$-effect}
In anisotropic turbulence under the influence of rotation non-zero
off-diagonal Reynolds stresses are generated according to
(R\"udiger \cite{Ruediger1989})
\begin{eqnarray}
Q_{ij} = \Lambda_{ijk} \Omega_k
+\mbox{diffusive terms}.
\end{eqnarray}
Considering the computational domain as a small rectangular part of a
sphere, we note that $Q_{xy}$ describes horizontal (latitudinal), and
$Q_{yz}$ vertical (radial) transport of angular momentum. In the
present study the vertical ($z$) direction is taken to be the
preferred one.  Furthermore, we use $f_1 = 0.2$ and $f_0 = 0$ which
results in turbulence that is
dominated by the $z$-component which, bearing in mind the geometry of
the system, is also the case for convection. The ratio of vertical to
horizontal turbulence intensity is $\overline{u_z^2}/\overline{u_{\rm
    H}^2} \approx 2.2$, where $\overline{u_{\rm H}^2} =
(\overline{u_x^2} + \overline{u_y^2})/2$.
A resolution of $32^3$ was used in the present calculations with ${\rm
  Re} \approx 7$. The Reynolds number dependence of the
$\Lambda$-effect is weak if ${\rm Re}$ is sufficiently large (not
shown). This issue will be studied further in a future publication.

\begin{figure*}
\resizebox{0.935\hsize}{!}
{\includegraphics{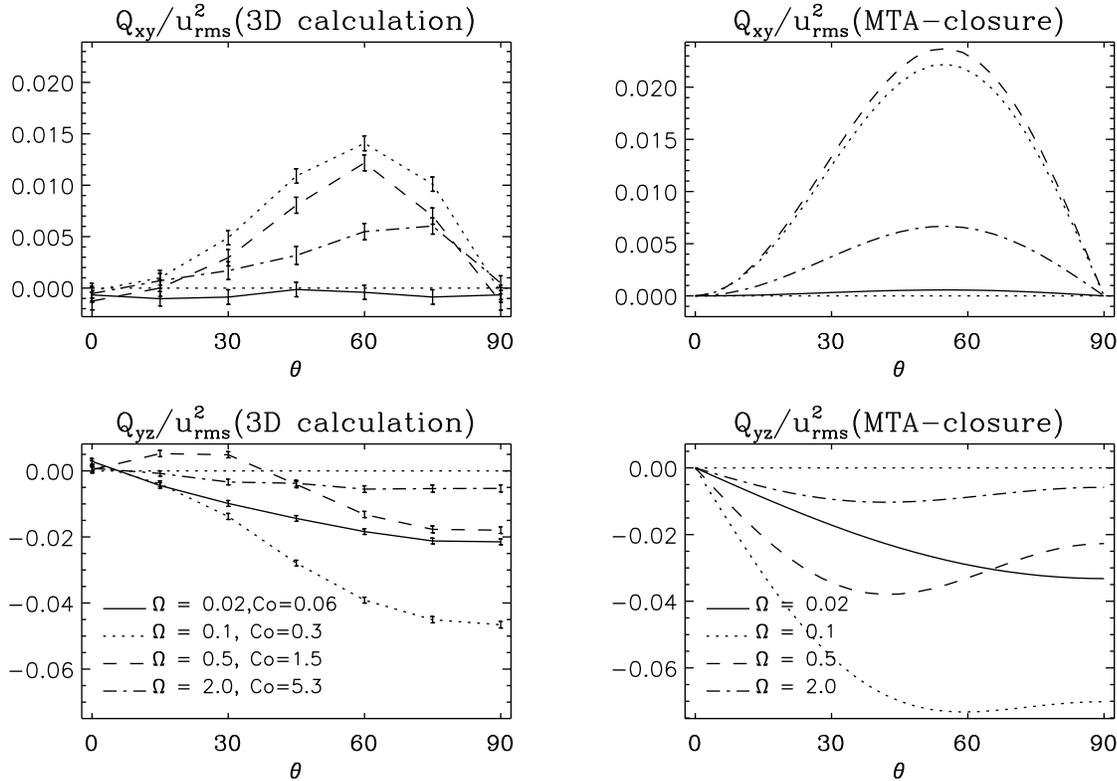}}
\vspace{-0.4cm}
\caption{The non-diffusive Reynolds stresses (i.e. $\Lambda$-effect)
  $Q_{xy}$ (upper panels) and $Q_{yz}$ (lower panels) from direct 3D
  simulations (left panels) and the MTA-closure (right panels) as
  functions of latitude and rotation rate. The error bars are defined
  similarly as in Fig.~\ref{fig:nut}.}
\label{fig:lambda}
\end{figure*}

The left hand panels of Fig.~\ref{fig:lambda} show the results from
direct 3D calculations. The rotational influence is quantified by
the Coriolis number
\begin{eqnarray}
{\rm Co} = 2\, \Omega_0 (u_{\rm rms} k_{\rm f})^{-1}\;.
\end{eqnarray}
Note that in comparison to the commonly used definition in convection
simulations, our Coriolis numbers are smaller by a factor of $2\pi$. The main
feature of the horizontal stress is that it is always positive and
peaks around 30 degrees latitude for all Coriolis numbers studied so
far. This result is in stark contrast to the convection calculations
where $Q_{xy}$ peaks always very near the equator (e.g. Chan
\cite{Chan2001}; K\"apyl\"a et al.\ \cite{Kaepylaeea2004}; Hupfer et
al.\ \cite{Hupferea2005}).

The vertical stress is predominantly negative with a maximum at the
equator. Although there seems to be a regime in which the vertical
stress is positive for intermediate rotation rates, the sign remains
negative for the most rapid rotation cases studied so far. This
feature is also at odds with convection simulations, which exhibit
positive $Q_{yz}$ for rapid enough rotation (K\"apyl\"a et
al.\ \cite{Kaepylaeea2004}; Chan\ 2007, private communication).

The MTA-model results are shown in the right panels of
Fig.~\ref{fig:lambda}. Qualitatively the results match the numerical
simulations rather well. The sign and latitude distribution of
$Q_{xy}$ is reproduced fairly well. The amplitude, however,
is clearly too large. Similar conclusions can be drawn from the
results for $Q_{yz}$, although there the latitude distribution from
the closure model tends to show a persistent maximum at mid-latitudes
which is not observed in the 3D simulations. Also the amplitude is too
large by almost a factor of two.

A possible explanation of the discrepancies between the numerical
results and the closure model is that the latter does not take
isotropizing effects properly into account. This can be due to an
improper treatment of the pressure terms in the closure model.

\section{Conclusions}
\label{sec:conclusions}
Shear flow turbulence simulations show that the ratio of turbulent to
molecular viscosity increases linearly up to ${\rm Re} \approx
30$ with $\nu_{\rm t}/\nu\approx1.5\,\mbox{Re}$.
For the largest Reynolds number the scaling seems somewhat
shallower but the present data is not yet sufficient to substantiate
this.
The $\Lambda$-effect from homogeneous, anisotropic turbulence does not
exhibit the puzzling features found in convection simulations. Further
study is required in order to understand which of the neglected
physics is responsible for the lack of these features.

The MTA-closure is able to reproduce many of the qualitative aspects
of the simulation results including a maximum of the horizontal stress
at about $30^\circ$ latitude, with its largest value for
$\mbox{Co}\approx0.5$.
In the model, the vertical stress can have a maximum away from the
equator for $\mbox{Co}\ga0.2$, which is not seen in the simulations.
Nevertheless, both simulations and model have the largest vertical
stress for $\mbox{Co}\approx0.3$.
However, the model generally overestimates the
magnitudes of the stresses. More detailed analysis of the simulation
and closure results will be presented in a future publication.

\acknowledgements{PJK acknowledges the financial support from the
  Helsingin Sanomat foundation. The simulations were performed with
  the computers hosted by CSC in Espoo, Finland. CSC is the Finnish IT
  center for science and is financed by the Ministry of Education.}

\end{document}